\documentclass[12pt]{article}

\usepackage{amssymb}
\usepackage{amsmath}
\usepackage{graphicx}

\begin{document}

\def\beq{\begin{eqnarray}}    
\def\eeq{\end{eqnarray}}      

\def\ln{\,\mbox{ln}\,}                  
\def\tr{\,\mbox{tr}\,}                  
\def\Tr{\,\mbox{Tr}\,}                  
\def\det{\,\mbox{det}\,}                
\def\Det{\,\mbox{Det}\,}                
\def\Res{\,\mbox{Res}\,}                
\renewcommand{\Re}{\,\mbox{Re}\,}       
\renewcommand{\Im}{\,\mbox{Im}\,}       
\def\lap{\Delta}                        

\def\al{\alpha}
\def\be{\beta}
\def\ch{\chi}
\def\ga{\gamma}
\def\de{\delta}
\def\ep{\varepsilon}
\def\ze{\zeta}
\def\io{\iota}
\def\ka{\kappa}
\def\la{\lambda}
\def\na{\nabla}
\def\pa{\partial}
\def\ro{\varrho}
\def\si{\sigma}
\def\om{\omega}
\def\ph{\varphi}
\def\ta{\tau}
\def\th{\theta}
\def\te{\vartheta}
\def\up{\upsilon}
\def\Ga{\Gamma}
\def\De{\Delta}
\def\La{\Lambda}
\def\Si{\Sigma}
\def\Om{\Omega}
\def\Te{\Theta}
\def\Th{\Theta}
\def\Up{\Upsilon}


\centerline{\large
\sc Effective Potential in Curved Space and 
Cut-Off Regularizations}
\vskip 8mm

\centerline{Fl\'{a}via Sobreira$^{(a)}$, 
\ \ \ Baltazar J. Ribeiro 
\ and \  
Ilya L. Shapiro$^{(b)}$} 
\vskip 6mm

(a) IFT-UNESP, S\~{a}o Paulo, SP, Brazil

(b) Departamento de Fisica, UFJF, Juiz de Fora, MG, Brazil
\vskip 12mm

\begin{quotation}
\noindent
{\bf Abstract.} \ \  
We consider derivation of the effective potential for a scalar 
field in curved space-time within the physical regularization 
scheme, using two sorts of covariant cut-off regularizations. 
The first one is based on the local momentum representation 
and Riemann normal coordinates and the second is operatorial 
regularization, based on the Fock-Schwinger-DeWitt proper-time 
representation. We show, on the example of a self-interacting 
scalar field, that these two methods produce equal results for 
divergences, but the first one gives more detailed information 
about the finite part. Furthermore, we calculate the contribution 
from a massive fermion loop and discuss renormalization group 
equations and their interpretation for the multi-mass theories. 
\vskip 4mm

\noindent
{\sl Keywords:} \ 
Effective potential, Curved space, Normal coordinates,
Renormalization Group, Renormalization schemes.
\vskip 4mm

\noindent
{\sl PACS:} \
04.62.+v, \
04.60.Gw, \
11.15.Kc.
\vskip 4mm

\noindent
{\sl AMS:} \
81T15, \ 
81T17, \ 
81T20. \ 
\end{quotation}
\vskip 8mm

\section{Introduction}

Recently there was a growing interest in the more physical 
regularization and renormalization schemes in curved space-time. 
In particular, one can mention the papers on deriving the 
energy-momentum tensor of vacuum in momentum cut-off regularization 
\cite{DeWitt-75,Akhmedov,PibRat,nob,Magg1,Bilic,Magg2} from one 
side and intensive discussions of physical interpretation of 
renormalization group from another one 
\cite{RW,Reuter1,cosm,nova,babic,CCfit,Gruni,DCCR,RotCurves,Stefan}. 

One of the outputs of the works on the cut-off approach is that 
this regularization may produce an explicit breaking of the 
local Lorentz invariance \cite{dewitt-75} and also of 
general covariance. Therefore, it would be interesting to have 
an example of the cut-off-based calculations which preserve both 
symmetries explicitly. 

The effective potential of a scalar field in curved space-time
have been studied in a number of papers starting from 
\cite{{GribMost,LapRub,Shore,buod-84}} (see \cite{book} for 
further references). In particular, a very general expression 
for such an effective potential has been obtained in 
\cite{buod-84} via the renormalization group method. Indeed, 
this means the Minimal Subtraction scheme of renormalization, 
when the effect of masses of the quantum fields are either 
ignored or taken into account through the heuristic method. 
An additional motivation for a more physical renormalization 
and regularization schemes comes from inflationary side. In 
the recent paper \cite{BezSha-2008} it was shown (see also 
previous works \cite{BoryaSetc} in this direction) that the 
Higgs-based inflation, originally invented by A. Guth \cite{Guth}, 
can be consistent with known observational tests if assuming 
that the Higgs field $H$ couples non-minimally to
scalar curvature. Let us remark that the corresponding  
term \ $\xi RH^* H$ \ is requested in order to 
make Standard Model of elementary particles multiplicatively 
renormalizable in curved space-time \cite{book}. 
The value of $\xi$ should be of the order of $10^4-10^5$, 
but this does not pose a problem, because the dimensional 
quantity $\left|\xi R\right|$ does not exceed the square 
of the Higgs mass. The great difference between the 
Higgs-based and inflaton-based inflationary models is 
that the Higgs field probably does exist. Therefore, the 
model of \cite{Guth,BezSha-2008} should be considered as 
the first candidate to describe an inflationary paradigm 
\cite{PoImpo}. According to the further works on Higgs 
inflation \cite{BKS-2008} and \cite{BezSha-2009}
(see also \cite{SimHerWil} and references therein), the 
renormalization group - based quantum correction to the Higgs 
potential play an essential role in this inflationary model, 
such that taking them into account leads to important 
restrictions for the Higgs mass. 
This result was essentially based on the well-known 
renormalization group derivation of effective potential
in curved space-time, completely equivalent to the one 
which was first performed in \cite{buod-84,book}), and 
concerns both one- and two-loop contributions. However, 
as far as this derivation is based on the Minimal Subtraction 
scheme of renormalization, it would be interesting to verify 
what is the effects of the masses of quantum fields by direct 
calculation. 

In order to address the issues of covariant cut-off and 
of the effect of masses on the Higgs potential in curved 
space, we perform direct calculation of the one-loop 
effective potential of a scalar field in curved space-time.
We consider two sorts of covariant cut-off regularizations. 
The first one is based on the local momentum representation,
which is due to the use of Riemann normal coordinates, 
and the second is the so-called operatorial regularization, 
based on the Fock-Schwinger-DeWitt proper-time representation. 
It was demonstrated recently in \cite{Liao} that these two 
types of regularizations give equivalent results in flat 
space-time. In view of this, our calculations can be seen as 
an extension of the same statement to a curved space. 

The paper is organized as follows. In the next section we
perform calculation for a self-interacting scalar field
through the cut-off regularization in the local momentum 
representation. In Sect. 3 we consider a technically simpler 
scheme of operatorial regularization cut-off. In Sect. 4 we 
extend the previous results to the fermion contributions. 
In Sect. 5 the $\mu$-dependence and renormalization group 
equations for the parameters of the theory are discussed. 
Finally, in Sect. 6 we draw our conclusions. 

\section{Covariant momentum cut-off calculation}

The effective potential is defined as the zero-order term in 
the derivative expansion of the effective action of a mean 
scalar field, 
\beq
\Ga[\ph,\,g_{\mu\nu}]\,=\,
\int d^4x\,\sqrt{-g}\Big\{
-\,V_{eff}(\ph) + \frac12\,Z(\ph)
\,g^{\mu\nu}\,\pa_\mu\ph\,\pa_\nu\ph
+ ... \,\Big\}\,.
\label{EffPot}
\eeq
The calculation of $\,V_{eff}(\ph)\,$ can be performed 
for constant $\ph$, in different theories with different 
content of quantum fields. In this paper we consider two 
examples, namely self-interacting scalar field and also 
fermion field with Yukawa coupling to the background
scalar, both in curved space-time. 

Our starting point will be the action of a real scalar 
field
\beq
S_0=\int d^4x\,\sqrt{-g}\Big\{\,
\frac12\,g^{\mu\nu}\,\pa_\mu\ph\,\pa_\nu\ph
- \frac12\,(m^2 - \xi\,R)\ph^2
- V(\ph)\,\Big\}\,,
\label{scalar}
\eeq
where $V(\ph) + m^2\ph^2$ is the minimal potential term
and $\,\xi R\ph^2$ is the non-minimal addition, which is 
necessary for formulating renormalizable theory in curved 
space-time.
In flat space $R=0$ and hence the non-minimal term vanish. 
Our purpose is to derive one-loop correction to Eq. 
(\ref{scalar}) in the constant scalar case. We perform 
calculations in four space-time dimensions. Hence we are 
mainly interested in the renormalizable case $V=f\ph^4/4\!\,$.
However in this section we shall use general notation 
$V(\ph)$, as being more compact and general. Let us emphasize 
that the theory of scalar field (\ref{scalar}) is renormalizable 
in the framework of semiclassical gravity \cite{book}. In this 
approach the metric is not quantized and represents a classical 
background for the quantum matter (in our case scalar) fields. 
The consistency and status of semiclassical approach has been 
recently discussed in \cite{PoImpo}.

In what follows we briefly consider the flat case first. One 
can see, e.g., \cite{VolTer} for a very pedagogical exposition 
with full details, despite there is some difference with our 
method. Recently, a similar calculation in flat 
space-time has been performed in \cite{Granada}, for a model of 
two scalar fields coupled to massless fermions. Since the 
main target of this work was an application to cosmology, 
it would be interesting to extend the result by taking 
curvature into account. 

At the second stage of the work, we will take care about 
linear in curvature corrections. We stop at the first order 
because it is sufficient for our purposes and because 
calculations become too cumbersome in the next-order 
approximation. However, the normal coordinate 
method enables one, in principle, to perform calculations to 
any given order in curvature tensor and its derivatives and 
also can be helpful to evaluate higher loops contributions. 

\subsection{Flat space calculation}

The result for the flat space is pretty well known 
\cite{ColeWein}. The derivation for the massive case
can be found, e.g., in the text-book \cite{VolTer}, where
it was obtained via Feynman diagrams. We can also arrive 
at the same result via the path integral functional 
methods. The starting point is the following expression: 
\beq 
V_{eff}(\ph)\,=\,m^2\ph^2 \,+\, V(\ph)\,+\,\bar{V}_0(\ph)\,,
\label{eff} 
\eeq
where 
\beq 
\bar{V}_0(\ph)\,=\,\frac{1}{2}\,\Tr\ln S_2(\ph) 
\,-\,\frac{1}{2}\,\Tr\ln S_2(\ph=0)\,,
\label{TrLn} 
\eeq
where $S_2(\ph)$ is the bilinear form of the classical action
in the background-field formalism \cite{BFM}. 
The last term in (\ref{TrLn}) can be seen as normalization of 
a functional integral. This term arises naturally through the 
diagrammatic representation of effective potential (see, e.g.,
\cite{VolTer}). In curved space-time the second term gets 
dependent on the metric and hence become relevant. Here and below 
we omit an infinite volume factor. Let us note that the one-loop
contribution (\ref{TrLn}) represents a quantum correction
to the complete expression $V(\ph) + m^2\,\ph^2/2$ and not 
just for \ $V(\ph)$. The same notations will be used in 
what follows. 

By introducing four-dimensional momentum cut-off $\Omega$,
we arrive at the result\footnote{In all momentum integrals 
we assume that the Euclidean rotation is performed.}
\beq 
\bar{V}_0(\ph,\eta_{\mu\nu})
&=& \frac{1}{32\pi^2}\,\int\limits_0^\Omega
k^2\,dk^2\,
\ln\Big(\frac{k^2+m^2+V^{\prime\prime}}{k^2+m^2}\Big)\,.
\label{k} 
\eeq
After taking this intergal we obtain 
\beq 
\bar{V}_0(\ph,\eta_{\mu\nu})
&=& \bar{V}_0 \,=\, \bar{V}_0^{div}+\bar{V}_0^{fin}\,,
\label{potya} 
\\
\bar{V}_0^{div}
&=&   \frac{1}{32\pi^2}\,\Big\{
\Om^2V^{\prime\prime} 
- \frac12\,\big(m^2 + V^{\prime\prime} \big)^2
\ln \frac{\Om^2}{m^2}\Big\}\,,
\label{potya-div} 
\\
\bar{V}_0^{fin} 
&=&    \frac{1}{32\pi^2}\,\Big\{
\frac12\,\big(m^2+V^{\prime\prime}\big)^2\,
\ln \big(1+\frac{V^{\prime\prime}}{m^2}\big)  
-\frac14\,\big(m^2+V^{\prime\prime}\big)^2
\Big\}\,.
\label{potya-fin} 
\eeq
In the last expressions we have included the $\ph$-independent 
$m^4$-type terms, which are indeed part of the second term in 
(\ref{TrLn}). 
The naive quantum contribution (\ref{potya}) must be supplemented
by an appropriate local counterterm, which we choose in the form 
\footnote{For the sake of convenience we have included into $\De {V}$ 
the finite term, this can be easily compensated by changing $\mu$.}
\beq 
\De {V}_0 &=& 
\frac{1}{32\pi^2}\,\Big\{
- \Om^2V^{\prime\prime} 
+ \frac12\,\big(m^2 + V^{\prime\prime} \big)^2\,
\ln \frac{\Om^2}{\mu^2}
+ \frac14\,\big(m^2+V^{\prime\prime}\big)^2\Big\}\,.
\label{contr-potya} 
\eeq
As a result we eliminate both quadratic and logarithmic 
divergences and arrive at the simple form of renormalized
effective potential 
\beq 
V_{eff,\,0}^{ren}(\eta_{\mu\nu},\,\ph) 
&=& \,m^2\ph^2 \,+\, V\,+\,\bar{V}_0\,+\,\De {V}_0
\nonumber
\\
&=& m^2\ph^2 \,+\, V\,+\,
\frac{1}{64\pi^2}\,
\big(m^2 + V^{\prime\prime} \big)^2
\ln \Big(\frac{m^2 + V^{\prime\prime}}{\mu^2}\Big)\,.
\label{ren-potya} 
\eeq

Looking at the counterterms (\ref{contr-potya}) it is easy to see 
that the renormalizable theory is the one which has \  
$V(\ph) = const \times \ph^4$. The reason is that for this 
potential the counterterms have the same form as the classical 
potential with an additional cosmological constant. At the next 
stage we will see that the same feature holds in curved space 
if the non-minimal term $\xi R\ph^2$ is introduced. 

\subsection{Riemann normal coordinates}

Riemann normal coordinates represent a useful tool for deriving 
local quantities, such as divergences or effective potential. 
These coordinates are based on the geodesic lines which link 
some fixed point $P^\prime(x^{\mu\prime})$ with other points. 
We can always assume that $g_{\mu\nu}(P^\prime) = \eta_{\mu\nu}$. 
One can fix the initial conditions for the geodesic lines in 
such a way that the metric in the point $P(x^\mu)$ becomes a 
Taylor series in the deviation $y^\mu = x^{\mu\prime}-x^\mu$. 
The coefficients of such an expansion are curvature tensor, 
its contractions and covariant derivatives at the point 
$P^\prime$. In the present work we will be interested only 
in the first order in curvature terms, and therefore all 
expansions will be taken in linear approximation. 

For instance, for the metric tensor we meet \cite{Petrov}
\beq
g_{\al\be}(x) = g_{\al\be}(x^{\prime})
-\frac13\,R_{\al\mu\be\nu}(x^\prime)\,y^\mu \,y^\nu\,.
\label{expansion 2}
\eeq

The bilinear operator of the action (\ref{scalar}) is 
\beq
- \,{\hat H} &=& 
- \frac{1}{\sqrt{-g}}\,
\frac{\de^2 S_0}{\de\ph(x)\,\de\ph(x^\prime)}
\,=\,  \Box + m^2 - \xi R + V^{\prime\prime}
\nonumber
\\
&=&
\eta^{\mu\nu}\pa_\mu\pa_\nu
+ \frac13\,{{{R^\mu}_\al}^\nu}_\be\,y^\al y^\be \pa_\mu\pa_\nu
-\frac23\,R^\al_\be\,y^\be \pa_\al \,+\,m^2\,-\,\xi R
 + V^{\prime\prime} \,.
\label{scalar 4}
\eeq
Of course, the term $\,-\xi R\,$ must be 
also expanded, but as far as we keep only first order in 
curvature, this is not relevant. 

The main advantage of the local momentum representation is 
that all calculations can be performed in flat space-time
(but with modified elements of Feynman technique) and the
result for some local quantity can be always presented in 
a covariant way. For instance, the equation for the 
propagator of the scalar field has the form 
\beq
{\hat H}\,G(x,x^\prime)\,=\,
-\,{g}^{1/4}(x^\prime)\,\de(x,x^\prime){g}^{1/4}(x)\,.
\label{scalar 5}
\eeq
It proves better to work with the modified propagator 
\cite{BunPar} $\bar{G}(x,x^\prime)$, where 
\beq
{\hat H}\,\bar{G}(x,x^\prime)\,=\,-\,\de(x,x^\prime)\,.
\label{scalar 6}
\eeq
It is important for us that the {\it r.h.s.} of the last 
relation does not depend on the metric, because we are going 
to use the relation $\Tr\ln {\hat H}=-\Tr\ln G(x,x^\prime)$ 
to obtain the dependence on curvature. 

The explicit form of $\bar{G}(x,x^\prime)$ is known for a 
long time \cite{BunPar} for the free $V^{\prime\prime}=0$
case. As far as $V^{\prime\prime}=const$, we can simply 
replace $m^2$ by ${\tilde m}^2 = m^2 + V^{\prime\prime}$ 
and obtain, in the linear in curvature approximation,  
\beq
{\bar G}(y)\,=\,\int \frac{d^4k}{(2\pi)^4}\,e^{iky}\,
\Big[\,
\frac{1}{k^2+{\tilde m}^2}
\,-\,\frac{(\xi-1/6)\, R}{(k^2+{\tilde m}^2)^2}
\Big]\,.
\label{scalar 7}
\eeq
Now it is a simple exercise to expand 
$\Tr\ln {\hat H}=-\Tr\ln G(x,x^\prime)$ up to the first 
order in the scalar curvature. We define
$$
{\hat H}={\hat H}_0+{\hat H}_1R + {\cal O}(R^2)\,,
\qquad  
{\bar G}={\bar G}_0+{\bar G}_1R + {\cal O}(R^2)
$$ 
and consider 
\beq
-\frac{1}{2}\,\Tr\ln {\bar G}(x,x^\prime) 
\,=\, \frac{1}{2}\,\Tr\ln \big( {\hat H}_0+{\hat H}_1R \big) 
\,=\, \frac{1}{2}\,\Tr\ln {\hat H}_0 
+ \frac{1}{2}\,\Tr \big( {\bar G}_0\,{\hat H}_1\,R \big)\,. 
\label{expand-1}
\eeq
The first term in the last expression has been calculated in 
the previous subsection, and the second one can be transformed
as follows:
\beq
\frac{1}{2}\,\Tr \big( {\bar G}_0\,{\hat H}_1\,R \big)
&=& -\,\int d^4x\,V_1\,R 
\,=\, \frac{1}{2}\,
\Tr \big[ {\bar G}^{-1}_0(x^{\prime\prime},\,x^\prime)
\,{\bar G}_1(x^\prime,\,x) \big]\,R
\label{expand-2}
\\
&=& 
\frac{1}{2}\,
\int d^4x\int d^4x^\prime \,
\big[ {\bar G}^{-1}_0(x,\,x^\prime)
\,{\bar G}_1(x^\prime,\,x) \big]\,R
\nonumber
\\
&=& \frac{1}{2} \,\int d^4x  \int d^4x^\prime \,R \,
\int \frac{d^4 k}{(2\pi)^4} e^{ik(x-x^\prime)}
\int \frac{d^4 p}{(2\pi)^4} e^{ip(x^\prime-x)}
\, {\bar G}^{-1}_0(k)\,{\bar G}_1(p)
\nonumber
\\
&=& \frac{1}{2}\,\int d^4x\,R\,
\int \frac{d^4 k}{(2\pi)^4} 
\, {\bar G}^{-1}_0(k)\,{\bar G}_1(-k)\,. 
\nonumber
\eeq
The last integration is trivial due to a simple form of 
${\bar G}_0(k)$ and ${\bar G}_1(k)={\bar G}_1(-k)$ in 
(\ref{scalar 7}), the final result reads 
\beq
\bar{V}(\ph,\,g_{\mu\nu})
&=& \bar{V}_0 \,+\,\bar{V}_1\,R\,\,,
\qquad 
\bar{V}_1\,=\,\bar{V}^{div}_1\,+\,\bar{V}_1^{fin}\,,
\label{scalar 8}
\\
\bar{V}^{div}_1
&=&  \frac{1}{2(4\pi)^2}\,\Big(\xi-\frac16\Big)\, 
\Big\{\,-\,\Om^2\,+\,\big(m^2+V^{\prime\prime}\big)
\,\,\ln \frac{\Om^2}{m^2}\Big\}\,,
\label{scalar 8 div}
\\
\bar{V}_1^{fin}&=& -\,
 \frac{1}{2(4\pi)^2}\,\Big(\xi-\frac16\Big)\, 
\big(m^2+V^{\prime\prime}\big)
\ln \Big(\frac{m^2+V^{\prime\prime}}{m^2}\Big)\,.
\label{scalar 8 fin}
\eeq

Similar to the flat space case, the potential must be modified 
by adding a counterterm
\beq
\De {V}_1
&=&  \frac{1}{2(4\pi)^2}\,\Big(\xi-\frac16\Big)\, 
\Big\{\Om^2\,-\,\big(m^2+V^{\prime\prime}\big)
\,\,\ln \frac{\Om^2}{\mu^2}\Big\}\,,
\label{potya-contr-R}
\eeq
as a result one eliminates quadratic and logarithmic 
divergences and arrive at the renormalized expression
\beq
V_{eff,\,1}^{ren}(g_{\mu\nu},\,\ph) 
&=& -\,\xi \ph^2
\,-\, \frac{1}{2(4\pi)^2}\,\Big(\xi-\frac16\Big)\, 
\big(m^2+V^{\prime\prime}\big)
\,\,\ln \Big(\frac{m^2+V^{\prime\prime}}{\mu^2}\Big) \,.
\label{potya-ren-R}
\eeq
Obviously, the renormalizable theory is the 
one which has the non-minimal term in the classical expression
(\ref{scalar}), without this term we can not deal with the 
corresponding counterterm (\ref{potya-contr-R}).

Making covariant generalization of the flat-space result 
(\ref{ren-potya}) and summing it up with (\ref{potya-ren-R}),
we arrive at the complete one-loop renormalized expression 
\beq 
V_{eff}^{ren}(g_{\mu\nu},\,\ph) 
&=& \rho_\La \,+\,\frac12\,(m^2-\xi R)\ph^2 \,+\, V
\label{REN-potya} 
\\
&+& 
\frac{\hbar}{2(4\pi)^2}\,
\Big[\frac12\,\big(m^2 + V^{\prime\prime} \big)^2
\,-\, \Big(\xi-\frac16\Big)\, R\,
\big(m^2+V^{\prime\prime}\big)\Big]\,
\ln \Big(\frac{m^2 + V^{\prime\prime}}{\mu^2}\Big)\,,
\nonumber
\eeq
where we restored the loop expansion parameter $\hbar$ at 
its place and also included the classical density of the 
cosmological constant term, $\rho_\La$, both for the sake 
of completeness. 

Let us note that the ambiguity related to $\mu$ can be eliminated 
by imposing renormalization conditions. Furthermore, $\mu$ cancels
automatically if we take into account the renormalization relations 
for the coupling $f$ and mass $m$ in the renormalizable case 
$V=f\ph^4/4\!$. This follows from the overall $\mu$-independence
of the effective action. However, in curved space-time 
the dependence on $\mu$ may be a useful tool for exploring different 
limits of effective action, such as the limit of short distances, 
the limit of strong scalar field or their combination 
\cite{buod-84,book} (see further references therein). 
Furthermore, as it was recently discussed in \cite{DCCR} the 
$\mu$-dependence can be an indication to the physical running 
of the cosmological constant. 

The numerical evaluation of the relative importance of the 
gravitational term in (\ref{REN-potya}) is strongly dependent 
on the mass $m$ of the field under discussion, on the value 
of $\xi$ and on the magnitude of curvature scalar in the 
given physical problem. It is easy to see that the relation 
between ``flat'' and ``curved'' terms in  (\ref{REN-potya})
is the same for classical and quantum one-loop terms. In the 
case when the scalar field is the Standard Model Higgs, we 
can assume the mass of the order of $100\,GeV$. 
The magnitude of $\xi$ which is needed for the Higgs 
inflation model of \cite{BezSha-2008} is about $4\cdot 10^4$. 
Then it is easy to see that the value of curvature when 
the gravitational term in (\ref{REN-potya}) become
dominating, is defined from the relation $\xi R = m^2$, 
hence the critical value is $R \propto 0.25\,GeV^2$. 
In the cosmological setting the corresponding value of 
the Hubble parameter is, therefore, $H \propto GeV$,
which is much greater than the phenomenologically 
acceptable value. From one side, this shows that the 
requested value of $\xi$ is not unnaturally large, 
because the dimensional product $\xi R$ remains small 
in at least most of the inflationary period. From another 
side, as it was discussed in \cite{BKS-2008,BezSha-2009} 
(see further references therein) the predictions of the 
theory are sufficiently sensible to the quantum corrections 
and this can lead to the constraints on the Higgs mass. 

\section{Operatorial cut-off regularization}

Another possibility is to implement the cut-off regularization in a
covariant manner via the Schwinger-DeWitt proper-time representation. 
Let us note that similar calculation for the massless case, using 
dimensional regularization, has been performed earlier in 
\cite{Ishikawa}.
 
The effective action can be written in the form (in Euclidean case)
\beq
{\bar \Ga}^{(1)} \,=\,
\frac{1}{2}\,\Tr\,\ln\,{\hat H}\,=\,
\frac{1}{2}\,\,\Tr\,\lim_{x^\prime \to x}
\int\limits_{1/L^2}^{\infty}\,\frac{ds}{s}\,e^{ -is\,{\hat H}}\,,
\label{e112}
\eeq
where $L$ is the cut-off parameter. Let us remember that the 
heat-kernel can be presented as \cite{BDW-65}
(see also \cite{bavi85,Avramidi-Thesis} and further 
references therein)
\beq
{\hat U}(x,x^\prime\,;s)\,=\,
e^{ -is\,{\hat H}} \de(x,x^\prime)\,=\,
{\hat U}_0(x,x^\prime\,;s)
\,\sum_{k=0}^{\infty}\,(is)^k\,{\hat a}_k(x,x^\prime)\,,
\label{dw}
\eeq
where 
\beq
{\hat U}_0(x,x^\prime\,;s)\,=\,
\frac{1}{(4\pi i)^{n/2}}\,
\frac{{\cal D}^{1/2}(x,x^\prime)}{s^{n/2}}\,
e^{\frac{i\si (x,x^\prime)}{2s} - im^2s} \,.
\label{e13}
\eeq
Here $\si (x,x^\prime)$ is the geodesic distance between
the two points $x$ and $x^\prime$. $\si (x,x^\prime)$
satisfies an identity $2\si=(\na \si)^2=\si^\mu\si_\mu$
and vanish in the coincidence limit $x^\prime \to x$.
${\cal D}(x,x^\prime)$ is the Van Vleck-Morett determinant
\beq
{\cal D}(x,x^\prime)\,=\,\det\Big[
-\,\frac{\pa^2 \si (x,x^\prime)}{\pa x^\mu\,\pa x^{\prime\nu}}\Big]\,,
\label{v1}
\eeq
which is a double tensor density, with respect to the both 
space-time arguments $x$ and $x^\prime$. 

Taking into account the mentioned features of the 
geodesic distance $\si (x,x^\prime)$, it is easy to see that 
the divergences are concentrated in the coincidence limits
of the first three Schwinger-DeWitt coefficients, namely 
\beq
{\bar \Ga}^{(1)}_{div} 
&=&
-\frac12\,\Tr\,\Big[ \frac{1}{2}\,a_0 \,L^4
\,+\,  a_1 \,L^2 
\,+\,  a_2 \, \ln \Big(\frac{L^2}{\mu^2}\Big) \Big]\,,
\label{divs}
\\
\nonumber
\mbox{where} \quad 
a_k &=& \lim_{x^\prime \to x} {\hat a}_k(x,x^\prime)\,.
\eeq 

The expressions for the $a_0,\,a_1$ and $a_2$ are well known
\cite{BDW-65} (see also Appendix B of \cite{frts-82} for the 
expressions with cut-off regularizations). 
In the scalar field case, for a constant background 
field we immediately obtain from (\ref{divs}) the expression
\beq
{\bar V}^{(1)}_{div} (scalar)
&=&
\frac{1}{2\,(4 \pi)^2}\,\Big\{ - \frac{L^4}{2}
\,+\, \Big[m^2 + V^{\prime\prime} - 
\Big(\xi-\frac16\Big)R\Big]\, L^2
\label{divs scal}
\\
&-&
\Big[\frac{1}{2}\,(m^2 + V^{\prime\prime})^2 
\,-\, \Big(m^2 + V^{\prime\prime}\Big)\Big(\xi-\frac16\Big)\,R\Big]
\,\ln \Big(\frac{L^2}{\mu^2}\Big) \Big\}\,,
\nonumber
\eeq 
where we disregarded the higher-curvature terms. The comparison 
between the divergences calculated with the two types of cut-off
shows that (\ref{divs scal}) is equivalent to the sum of 
(\ref{potya-div}) and (\ref{scalar 8 div}) if we identify 
the two cut-off parameters $\Om$ and $L$. 

It is possible now to make a comparison between the two cut-off 
schemes. The both give equivalent $\ph$-dependent divergent 
parts, however the local momentum cut-off method is capable to 
provide also a complete expressions for the finite part of the 
one-loop effective potential 
(\ref{potya-fin}) and (\ref{scalar 8 fin}). In the case of the 
proper-time cut-off scheme one can also arrive at the same 
renormalized expression $(\ref{REN-potya})$ through the 
renormalization-group approach \cite{buod-84}. However this 
requires an \ {\it ad hoc} \ identification of $\mu^2$ with 
$m^2 + V^{\prime\prime}$. 
At the same time such an identification arises quite naturally 
within the local momentum cut-off method, because in this case 
we can work directly with the finite part of the renormalized 
effective potential (\ref{REN-potya}). 

Some additional remark would be in order. A natural tentation 
would be calculate the effective potential directly by using
the method of summing up the Schwinger-DeWitt series \cite{bavi90}.
However this idea meets an obstacle when it is used to calculate 
static quantities such as quantum corrections to the cosmological 
constant \cite{apco}. The reason is that the final output of 
this approach is a form factor which is given by an algebraic 
function of D'Alembert operator $\Box$ (covariant Laplacian in 
Euclidean case) acting on generalized curvature. In the static 
case, $\Box$ acting on a constant gives zero and hence this method 
in its original form is not efficient. The same applies also to the 
scalar field potential, because according to Eq. (\ref{EffPot})
the derivatives of a scalar go to the next term of the 
expansion of effective action. It would be an interesting 
excercize to modify the Schwinger-DeWitt series in such a 
way that the derivation of finite quantum corrections to the
cosmological constant or to the potential of scalar field 
(these two are in fact closely related \cite{DCCR}) become
possible, but at the moment the perspective of such calculation 
looks unclear. At the same time one can perfectly calculate the 
effective potential in the form of expansion in curvature tensor
directly by using normal coordinates, as use did in the previous 
section\footnote{Unfortunately, this method is useless for 
the cosmological constant case.}. We can conclude that the 
two methods perfectly complement each other, because they 
enable one to identify $\mu^2$ as $m^2 + V^{\prime\prime}$
or other similar expression which shows up in other models. 
We will discuss an application of this idea in the next section. 

\section{Fermion contributions}

As a practical application of the equivalence between the 
two cut-off schemes, let us consider the contribution of the
fermion field with Yukawa interaction,
\beq
S_0=\int d^4x\,\sqrt{-g}\,\,i\, {\bar \psi} \,
\big(\,\gamma^\mu \nabla_\mu - im_f - ih\ph\,\big)\,\psi\,.
\label{spinor}
\eeq
As far as we are interested in the effective potential, the 
calculation can be done for a constant $\ph$ and hence we can 
denote 
\beq
{\tilde m} &=& m_f + h\ph\,.
\label{m tio}
\eeq
Taking the Grassmann parity into account, the object of 
our interest is \footnote{The operation of $\Tr$ is defined 
without taking statistics into account.}
\beq
{\bar \Ga}_f^{(1)}\big[\ph,\,g_{\mu\nu}\big]
\,=\, - \Tr \ln {\hat H}_f\,,\quad 
\mbox{where} \quad 
{\hat H}_f \,=\, i 
\big(\,\gamma^\mu \nabla_\mu - i{\tilde m}\,\big)\,.
\label{Hf}
\eeq
As far as the result is expected to be even in ${\tilde m}$
(see, e.g., \cite{GBP}), one can perform the transformation
\beq
\Tr \ln {\hat H}_f 
&=& \frac{1}{2}\,\Tr \ln \big(
{\hat H}_f\,{\hat H}_f^*\big)
\,,\quad 
\mbox{where} \quad 
{\hat H}_f^* \,=\, i 
\big(\,\gamma^\mu \nabla_\mu + i{\tilde m}\,\big)\,.
\label{Hf*}
\eeq
The last product can be cast into the form
\beq
{\hat H}_f\,{\hat H}_f^* \,=\,-\, 
\Big(\Box - \frac{1}{4}\,R + {\tilde m}^2\Big)\,.
\label{Box+}
\eeq
Using the proper-time method we arrive at the 
expression for divergences 
\beq
{\bar V}^{(1)}_{div} (fer)
= - \frac{2}{(4 \pi)^2}\,
\Big\{ - \,\frac{L^4}{2}
\,+\, \Big({\tilde m}^2  - \frac{1}{12}\,R\Big) L^2
\,-\, \frac{1}{2}\,\Big({\tilde m}^4
- \frac{1}{6}\,R {\tilde m}^2\Big)
\,\ln \Big(\frac{L^2}{\mu^2}\Big) \Big\}\,,
\label{fer div}
\eeq 
where we neglected the higher-curvature terms.

By using equivalence between the two cut-off 
schemes, we can easily write down the finite part of the 
renormalized one-loop contribution to the effective potential, 
namely
\beq
{\bar V}^{(1)}_{ren} (fer)
= - \frac{1}{(4 \pi)^2}\,
\Big({\tilde m}^4 - \frac{1}{6}\,R\,{\tilde m}^2\Big)
\,\ln \Big(\frac{{\tilde m}^2}{\mu^2}\Big) \,.
\label{fer fin}
\eeq 
If we compare this result to the one of the Minimal 
Subtraction scheme of renormalization, it is clear that 
the correct identification of \ $\mu^2$ \ is \
${\tilde m}^2 = (m_f + h\ph)^2$. 

\section{Renormalization group}

Let us come back to the scalar result (\ref{REN-potya}) and 
use it as an example of how the renormalization group equations 
for the parameters of the potential can be obtained. For this 
end we have to restrict our consideration by the renormalizable 
case $V=\la\ph^4/4!$, such that $V^{\prime\prime}=\la\ph^2/2$
and the counterterms $\,\De V = \De V_0 \,+\,\De V_1R$, with 
$\De V_0$ from (\ref{contr-potya}) and $\De V_1$ from 
(\ref{potya-contr-R}), have the same dependence on $\ph$ as 
the corresponding classical terms. 

In order to obtain the renormalization group equations 
for the parameters one has to assume that the renormalized 
effective potential is equal to the bare effective potential. 
This statement is an intrinsic feature of the effective action 
which can be easily proved in a general form (see, e.g., 
\cite{book}). For the finite part of effective potential this 
means that the apparent $\mu$-dependence of the 
renormalized effective potential (\ref{REN-potya}) must be 
compensated by the $\mu$-dependence of the independent 
parameters of the theory, namely $\la$, $m$ and $\rho_\La$. 
Therefore one can easily find  $\la(\mu)$, $m(\mu)$ and 
$\rho_\La(\mu)$ directly from (\ref{REN-potya}). We leave this 
calculation as an exercise for the interested reader and instead 
will obtain the corresponding $\be$-functions from the infinite 
renormalization of the classical action, similar as it is done 
in the MS-scheme and dimensional regularization \cite{book}.

The classical (extended by mass and non-minimal terms) potential,
with the added counterterm, form the renormalized classical 
potential, which should be equal to the bare one, 
hence\footnote{Here we mark bare parameters by the subscript 
$(0)$. In the simples case of purely scalar theory one
does not need to renormalize the field $\ph$, but in general 
case it is not so, of course.}
\beq
\rho_\La + (m^2-\xi R)\ph^2 + \frac{\la\ph^4}{4!}
+ \De {V}_0 + R\De {V}_1
= \rho_{\La(0)} 
+ \big[m_{(0)}^2-\xi_{(0)} R\big]\ph^2 
+ \frac{\la_{(0)}\ph^4}{4!}\,.
\label{Rena}
\eeq
The {\it l.h.s.} of this relation does depend on $\mu$ 
explicitly and the  {\it r.h.s.} does not. This condition
should be satisfied for all terms separately, because there 
are arbitrary quantities $\ph$ and $R$. Therefore, 
using (\ref{contr-potya}) and (\ref {potya-contr-R}), 
we arrive at the equations
\beq
\rho_{\La(0)}  &=& \rho_\La 
+ \frac{m^4}{2(4\pi)^2}\,\ln \frac{\Om^2}{\mu^2}\,,
\nonumber
\\
m_{(0)}^2 &=& 
m^2 + \frac{\la\, m^2}{2(4\pi)^2}\,\ln \frac{\Om^2}{\mu^2}\,,
\nonumber
\\
\xi_{(0)} &=& \xi + \frac{\la}{2(4\pi)^2}
\,\Big(\xi-\frac16\Big)\,\ln \frac{\Om^2}{\mu^2}\,,
\nonumber
\\
\la_{(0)} &=& \la + \frac{4!\,\la^2}{16\,(4\pi)^2}
\,\ln \frac{\Om^2}{\mu^2}\,.
\nonumber
\label{bare-R}
\eeq
At this stage we can apply the conventional wisdom to take 
derivatives $\mu \frac{d}{d\mu}$ of the bare quantities \
$\rho^{(0)}_\La$, $m_{(0)}^2$, $\xi_{(0)}$ and $\la_{(0)}$ \
and set them to zero. As a result we arrive at the following 
$\be$-functions:  
\beq
\mu\,\frac{d\rho_\La }{d\mu}  &=& \frac{m^4}{2(4\pi)^2}\,,
\qquad \rho_\La(\mu_0) = \rho_{\La,0}\,;
\nonumber
\\
\mu\,\frac{dm^2}{d\mu}  &=& \frac{\la}{(4\pi)^2}\, m^2\,,
\qquad m^2(\mu_0) = m^2_0\,;
\nonumber
\\
\mu\,\frac{d\xi}{d\mu}  &=& 
\frac{\la}{(4\pi)^2} \,\Big(\xi-\frac16\Big)\,,
\qquad \xi(\mu_0) = \xi_0\,;
\nonumber
\\
\mu\,\frac{d\la}{d\mu}  &=& 
\frac{3\,\la^2}{(4\pi)^2}\,.
\qquad \la(\mu_0) = \la_0\,,
\label{betas}
\eeq
where the initial points of the renormalization group 
trajectories is defined at some reference value (scale) $\mu_0$.
The solution of these equations is well-known, e.g., in the 
leading-log approximation we have 
$$
\la(\mu)
\,=\,\la_0 + \frac{3\,\hbar\,\la_0^2}{(4\pi)^2}\,\ln (\mu/\mu_0)\,,
$$
where we restored $\hbar$ for further convenience\footnote{We note 
that the solution in the momentum-subtraction scheme is much less 
simple, see \cite{Bexi}.}

It is easy to check that if we replace these solutions
into the renormalized effective potential (\ref{REN-potya}),
the dependence on $\mu$ completely disappears in the 
 ${\cal O}(\hbar)$-terms. Definitely, 
this does not mean that the effective potential becomes 
trivial, because the real content of quantum corrections is 
related to the dependence on $\ph$, which did not change 
under the procedure described above. 

We can conclude that the $\mu$-dependence is nothing but 
a useful tool for obtaining the dependence on  $\ph$, or 
on the derivatives of $\ph$ (or other mean field). 
This tool becomes especially 
important in those cases when the derivation of explicit 
dependence on the fields and their derivatives is not 
possible, as it was discussed recently in \cite{DCCR} for 
the case of external gravitational field. 

The last observation is that the relation between 
$\mu$-dependence and real effective potential may be 
rather nontrivial in a more complicated models. Consider, 
for example, a theory where the scalar field is coupled 
to different fermions with distinct masses. According 
to the result of the previous section, (\ref{fer fin}), 
there is no unique identification of $\mu$ in this 
case. Therefore, one need to be very careful when using
the renormalization group results for the massive theories, 
especially when different masses are present. 

\section{Conclusions}

We have performed an explicitly covariant calculation of 
effective potential in two types of cut-off regularizations. 
The divergences are identical within the two approaches, but 
the covariant local momentum cut-off has an advantage to  
provide also the finite part of effective potential for the 
massive case and, consequently, it indicates the physical 
interpretation for the renormalization group parameter $\mu$. 
It would be interesting to apply the same method to the 
derivation of the ``Energy-Momentum Tensor" of vacuum and,
in this way, resolve the amazing puzzle with non-covariant 
power-like divergences which were described recently in 
\cite{Akhmedov,Magg1,Bilic}. The work in this direction is 
in progress and the results of the calculations presented 
here are going to be useful in this respect. 

Another important conclusion of our work is the restricted 
sense of the renormalization group - based quantum corrections 
for the quantum theory of massive fields, especially if 
different masses are present. 

\section*{Acknowledgements}
Authors are very grateful to Neven Bilic, Silvije Domazet 
and Branco Guberina for carefully reading the paper and 
indicating some errors/misprints in signs. 
Authors are grateful to CAPES (B.R. and F.S.) for support, 
also to FAPEMIG (B.R. and I.Sh.), CNPq and ICTP (I.Sh.) for 
partial support of their work.


\end{document}